\documentclass[a4paper]{JHEP3}
\usepackage{epsfig}
\usepackage{graphicx}
\usepackage{amssymb}
\usepackage{latexsym}
\usepackage{amsmath}

\newcommand{\be}{\begin{equation}}
\newcommand{\ee}{\end{equation}}
\newcommand{\bea}{\begin{eqnarray}}
\newcommand{\eea}{\end{eqnarray}}

\title{Dirac and the Path Integral}
\author{N.D. Hari Dass \\  
Anuradha, 2710/2, Second Main Road\\V.V. Mohalla, Mysore 570002\\
Email: \email{dass@tifrh.res.in}}

\abstract{Through a very careful analysis of Dirac's 1932 paper on the Lagrangian in Quantum Mechanics as well as the second and third editions
of his classic book {\it The Principles of Quantum Mechanics}, I show that Dirac's contributions to the birth of the path-integral approach
to quantum mechanics is not restricted to just his seminal demonstration of how Lagrangians appear naturally in quantum mechanics, but that
Dirac should be credited for creating a path-integral  which I call {\it Dirac path-integral} which is far more general than Feynman's while
possessing all its desirable features. On top of it, the Dirac path-integral is fully compatible with the inevitable quantisation ambiguities, 
while the Feynman path-integral can never have that full consistency. In particular, I show that the claim by Feynman that for infinitesimal
time intervals, what Dirac thought were analogues were actually proportional can not be correct always.I have also shown the conection
between Dirac path-integrals and the Schr\"odinger equation. In particular, it is shown that each choice of Dirac path-integral yields a
{\it quantum Hamiltonian} that is generically different from what the Feynman path-integral gives, and that all of them have the same
{\it classical analogue}. Dirac's method of demonstrating the least action principle for classical
mechanics generalizes in a most straight-forward way to all the generalized path-integrals.\footnote {Dedicated to Herbert Jehle for playing 
the memorable match-maker.}
}


\begin{document}
\section{Introduction}
The path-integral formulation of quantum mechanics is clearly one of the most creative developments in modern physics. Called the 
{\it third formulation} of quantum mechanics by its creator Richard Feynman, it has impacted so many areas like polymer physics, quantum field theory, string theory and even cosmology \cite{kleinert}. It has paved the way for a mathematically exact mapping betweem quantum theories and
classical statistical mechanics, a spectacular outcome of which is {\it Lattice Gauge Theory} \cite{seiler}. This has produced very fundamental non-perturbative understanding of quantum field theories including {\it Quantum Chromodynamics}, currently the best bet theory of strong interactions.

It got such rave reviews that Freeman Dyson is supposed to have said "..this wonderful vision of the world as a woven texture of world lines in space and time, with everything moving freely" \cite{gleick}. While such hyperbole may inspire {\it physics for poets}, it is not of much
help for serious students of quantum theory! Dyson also went on to say "It was a unifying principle that would either explain everything or nothing". Actually, though it was a radically new formulation, it was completely equivalent to the older formulations by Heisenberg and Schr\"odinger, which were themselves equivalent to each other as elegantly and powerfully demonstrated by Dirac! In fact, Feynman himself beautifully
expressed it in his Nobel acceptance speech \cite{nobel}: " I wonder if anything can be learned from it. I doubt it. It is most striking that 
most of the ideas developed in the course of this research were not ultimately used in the final result. The path-integral formulation of 
Quantum Mechanics was useful for guessing at final expressions and at formulating the general theory of electrodynamics in new ways...
although, strictly it was not absolutely necessary."

Nevertheless, it was a new way, a radically new way, of looking at quantum mechanics notwithstanding the fact that it was completely equivalent
to the old ways. And, that in itself can usher important new developments, and even ones that may make a complete break from the past. The
Hamiltonian approach to classical mechanics is a case in point; though it is completely equivalent to the Newtonian formulation, it facilitated
the passage to statistical mehanics and even to quantum mechanics, an altogether different theory.

What led Feynman to his work that culminated in the path-integral formulation of quantum mechanics, what he called {\it the space-time
approach}\cite{feynpaper} was his hard quest for a way of quantising systems that could not be described by a Hamiltonian. An explicit
example of that was the absorber theory of electrodynamics that he developed with John Wheeler(see \cite{thesis}) . That theory was
described by a classical Lagrangean: 
\be
\label{eq:absorberL}
L\,=\,\frac{m}{2}\,{\dot x}^2\,+\,k^2\,{\dot x}(t)\,{\dot x}(t\,+\,T)
\ee
The Euler-Lagrangean equations of motion for this are:
\be
\label{eq:absorberEOM}
m\,{\ddot x}(t)\,+\,k^2\,{\ddot x}(t\,+\,T)\,k^2\,{\ddot x}(t\,-\,T)\,=\,0
\ee
This couples motions at different times and no Hamiltonian description is possible.

So Feynman was looking for a quantum mechanical description that involved only the {\it classical action}. As flamboyantly narrated by 
Feynman in his nobel lecture \cite{nobel} and other places \cite{joking}, and further propagated by his biographer
James Gleick \cite{gleick}, a chance meeting in 1941 with Herbert Jehle at a beer party put Feynman on the right track! Incidentally, 
I got to know Jehle quite well while working at the Max Planck Institute, Munich! He was a kind and sensitive gentleman despite all his 
sufferings. Answering Feynman whether he knew of any methods doing quantum mechanics directly in terms of actions, Jehle introduced him to 
Dirac's 1932 paper {\it The Lagrangean in Quantum Mechanics}\cite{diracpaper}. In that classic paper, Dirac had shown that the
quantum mechanical amplitude(also called the {\it propagator} some times) $ \langle\,q^\prime_t|q^\prime_T\rangle $,
where $|q^\prime_t\rangle$ are the eigenstates of the dynamical variable(say, position) $q_t$ at time t while $|q^\prime_T\rangle$
are the eigenstates of the same dynamical variable $q_T$ at a later time T, obtained from $q_t$ through Hamiltonian time evolution, was
the {\it quantum analogue} of $ e^{\frac{iS_{cl}}{\hbar}} $ with $S_{cl}$ the action(more precisely the Hamilton's principal function) of the 
classical system related to its classical Lagrangean by $ S_{cl}\,=\,\int_T^{t}\,L\,dt $. In particular, Dirac stated that when t and T are 
infinitesimally close,
\be
\label{eq:qanalogue}
\langle\,q^\prime_{t+\delta\,t}|q^\prime_t\rangle\,\sim\,\, e^{i\,\frac {L\,\delta\,t}{\hbar}}
\ee
where we have denoted what Dirac calls the {\it quantum analogue} by $\,\sim\,$. The importance Dirac attached to this result can be gauged
by his own words \cite{diracbook2}: {\it This result gives probably the most fundamental quantum analogue for the classical Lagrangian 
function.}
******

Feynman was perplexed by the word {\it analogue} and he is supposed to have asked Jehle the two famous questions:
{\bf What does he mean, they are analogous?}, and, {\bf What is the use of that?}. As already mentioned,
Feynman was looking for a more precise connection so he could carry out the quantisation of the systems he had in mind for which
there was no Hamiltonian desription, but only a classical action. We have explicitly discussed such a Lagrangean, given in 
eqn.(\ref{eq:absorberL}). He is supposed to have quipped to Jehle "Dirac must have meant they are
equal." To this, Jehle is supposed to have retorted "No, he doesn't mean they are equal".  

In what was clearly a major step, Feynman is supposed to have said " Let us see what happens if we make them equal" and proceeds to try 
and put them equal, albeit for infinitesimal time separations, and
for the case of $L\,=\,\frac{1}{2}\,M\,{\dot x}^2\,-\,V(x)$, shows that the Schr\"odinger equation can be recovered provided a 
{\it normalisation} factor is introduced into the equality, whereupon Feynman tells Jehle that Dirac meant they were {\it proportional}. To
this also Jehle is supposed to have responded "No, no. Dirac's idea had been strictly metaphorical; the englishman had not meant to
suggest that the approach was useful." \cite{gleick}. Jehle told Feynman that he, Feynman, had made an important discovery.

Expressed mathematically, what Feynman claimed was
\be
\label{eq:missinglink0}
\langle\,q_{t\,+\,\epsilon}^\prime|q_t^\prime\rangle\,=\,\frac{1}{A(\epsilon)}\,e^{i\,\frac{L\,\epsilon}{\hbar}}
\ee
when $\epsilon$ was small.

Two important points are worth stressing at this point. Feynman could have easily anticipated the normalisation factor, without any
detailed calculations. The l.h.s of eqn.(\ref{eq:qanalogue}) actually \emph{diverges} as $\epsilon\,\rightarrow\,0$ because of continuum 
normalisation (l.h.s approaches a Dirac delta function). The second is about Feynman's obtaining the Schr\"odinger equation; what he obtained
was the Schr\"odinger equation with the very specific quantum Hamiltonian
\be
\label{eq:feynHq}
{\hat H}_{feyn}\,=\,\frac{{\hat p}^2}{2\,m}\,+\,V({\hat x})
\ee
The significance of stressing this will become clear later.

At this point, we will club the three questions raised by Feynman into what we call \emph{Feynman's two questions}:\\
{\bf FQ1: What does he mean they are analogous? Did he mean they are equal? Did he mean they are proportional?}\\
{\bf FQ2: What is the use of that?}\\

It is worthwhile to quote verbatim Feynman's own narrative: "So, I thought I was finding out what Dirac meant, but, as a matter of fact, 
had made the discovery that what Dirac thought was analogous, was, in fact, equal. I had then, at least, the connection between the 
Lagrangian and quantum mechanics, but still with wave functions and infinitesimal times."\cite{nobel}

This crucial step, namely an equality in place of just an analogy, albeit for infinitesimal $\epsilon$, is what I would call the 
{\bf path-integral missing link} and I shall comment about it in depth later on. It is a missing 
link because once the short-time propagator is known, it can be iterated many times to get the path-integral representation (as both Dirac
and Feynman did). Next, Feynman iterates the infinitesimal result a large number of times to arrive at his famous path-integral formula. 
Again, to quote Feynman, "At last, I had succeeded in representing quantum mechanics directly in terms of the action $S_{cl}$. The
explicit path-integral representation obtained by Feynman is given below:
\be
\label{eq:feynpathint0}
K(x_t^\prime;x_T^\prime)\,=\,{\cal A}\,\int\int..\int\,\prod_{i=1}\,dx_i^\prime\,e^{i\frac{S}{\hbar}}
\ee
here $K$ is called the {\it Kernel} and has the same meaning as Dirac's $\langle\,q_t^\prime|Q_T^\prime\rangle$. The sequence 
$(x_t^\prime, x_1^\prime,x_2^\prime\,\ldots\,x_T^\prime$ is interpreted as a {\it path} with {\it fixed} endpoints $(x_t^\prime,x_T^\prime)$ 
and consequently eqn(\ref{eq:feynpathint0}) is construed as an
integral over 'all paths' connecting the fixed end-points. It is tacitly understood that one finaly passes to the limit when 
$\delta\,t\,=\,\frac{T\,-\,t}{N} \rightarrow 0$. Here N is the total number of subdivisions. Therefore, $\delta\,t\,\rightarrow\,0$ limit 
has to be realised by $N\,\rightarrow\,\infty$, keeping $T\,-\,t$ \emph{fixed}. ${\cal A}$ is a suitable normalisation, which is really infinite in the continuum limit but 
is discarded as harmless since it does not depend on the dynamical variables. A word of caution is in order here: it is important to consistently maintain either
the Schr\"odinger picture or the Heisenberg picture. Dirac uses the Heisenberg picture throughout. 

In his thesis \cite{thesis} and in his later paper \cite{feynpaper} Feynman referred to Dirac as follows: "Dirac had worked out the beginnings
of a least action theory"\cite{thesis}, "The formulation was suggested by some of Dirac's remarks..." \cite{feynpaper}, "The close analogy
between $\langle\,x^\prime|x\rangle_{\epsilon}$ and the quantity $e^{i\frac{S}{\hbar}}$ has been pointed out on several occasions by Dirac. In
fact, we now see that to sufficient approximations the two quantities may be taken to be proportional to each other. Dirac's remarks were the starting point of the present development. The points he makes concerning the passage to the classical limit $\hbar\,\rightarrow\,0$ are very
beautiful that I may perhaps be excused for briefly reviewing them here." \cite{feynpaper}. While Feynman reviews Dirac's ideas behind the principle
of least action, he does so within the framework of his own work. Unfortunately, that does not throw light on what {\it exactly} Dirac said, and
that is very important to understand how Dirac himself had already come to the path-integral concept. In Feynman's thesis, in the section on {\it Least Action in Quantum Mechanics}), he quotes verbatim(adding, in his own words "
A description of the proposed formulation of quantum mechanics
might best begin by recalling some remarks made by Dirac concerning
the analogue of the Lagrangian and the action in quantum
mechanics. These remarks bear so directly on what is to follow and
are so necessary for an understanding of it, that it is thought best to
quote them in full, even though it results in a rather long quotation.")
excerpts from the second edition of Dirac's famous book {\it The Principles of Quantum Mechanics} \cite{diracbook2}, but not from Dirac's paper.
There too Feynman  does not quote the parts where Dirac gives a precise mathematical basis for why he used the word \emph{analogues} . 
It is likely that the Feynman's first question resulted from  his not paying sufficient attention to these finer points. The flamboyant
folklore as well Feynman's own strong emphasis on the lagrangian and least action aspects have created an impression that Dirac's role in the 
development of the path-integrals was restricted only to these.

The purpose of this article is to point out that Dirac contributed a lot more. That his works clearly displayed the concept of a path-integral;
that he clearly demonstrated the meaning of the {\it principle of least action} of classical mechanics within a path-integral formulation; Dirac
essentially showed how quantum mechanical quantities could be computed with a classical calculus. I have carefully analysed Dirac's paper on the
Lagrangian in Quantum Mechanics of 1932, the second edition of his book of 1935 \cite{diracbook2} as well as the third edition of his book that came 
out in 1947 \cite{diracbook3}.
While subsequent editions of his book remained unchanged, as far as these issues were concerned, there are very important differences, both
of a conceptual as well as of styles of narration between the paper, and the second and third editions of the book.. All those differences will play an important role in the analysis presented here. They
also reveal Dirac's own struggles at a lucid and coherent formulation. Contrary to the editor L.M. Brown's footnote 13 \cite{thesis}, "
"Later editions contain very similar material regarding the
fundamental aspects to which Feynman refers", third edition is significantly different from second edition, which in itself is quite different
from Dirac's paper.

I will also present a detailed critique of the {\it missing link} issue, from both Dirac's and Feynman's perspectives. The fact that going 
from classical theory to its "quantised" version is necessarily non-unique, will play an important role in that analysis. The non-uniqueness
 has been explicitly pointed out by Feynman himself in his thesis: 
"The inverse problem, that of determining a quantum mechanical
description of a system whose classical mechanical behaviour is
known, may not be so easily solved. Indeed, the solution cannot be
expected to be unique." Unfortunately, he did not exactly say what he meant by \emph{the solution}! Did he mean the Schr\"odinger equation
that follows from path-integrals or did he mean the non-uniqueness of path-integrals to be associated with quantum systems with given
classical limits? Or did he mean the determination of quantum Hamiltonians with fixed classical limits? The claim here is that Dirac's
works, taken to their logical conclusions, resolve precisely these foundational issues.

Let me briefly describe Dirac's paper \cite{diracpaper} and the four editions of his famous book 
\cite{diracbook1,diracbook2,diracbook3,diracbook4,diracbook4r} to place these introductory remarks in perspective as also
to set the stage for the rest of this article. In the paper, Dirac provides a mathematically precise meaning to the analogy. This is 
based on a detailed comparison between classical and quantum {\it contact} transformations. This should have clearly answered Feynman's first 
question{\bf FQ1} in the negative, namely, that in the generic case the analogues are neither equal nor proportional to each other. 
Most importantly,
Dirac shows that the analogy holds for arbitrary values of $(T\,,t\,,\hbar)$. Dirac then introduces a sub-division of the interval [T,t] into a
large number of small intervals and examines the self-consistency of the analogy. In doing so, he arrives at a path-integral formula. He
shows that for $\hbar\,\rightarrow\,0$ self-consistency results in the action principle for classical mechanics. He then attempts to generalize
the discussion to arbitrary $\hbar$. He appears to get 
stuck in circularity. Nevertheless, a careful scrutiny of all his equations suggest a very neat resolution. That is to his definition, 
presented in eqn.(\ref{eq:Stilde}) below, to define a \emph{new} class of path-integrals, which we are calling {\it Dirac Path-integrals}.

In the second edition of the book \cite{diracbook2}, one finds several differences from the paper. Dirac drops the proof he gave in 
the paper for the analogy.
Instead, starting from Schr\"odinger equation, he shows that $\langle\,q_t^\prime|\,q_T^\prime\rangle$ also obey Schr\"odinger equation
both wrt to t and T, and shows their
correspondence with classical contact transformations. 
But Dirac states the issues concerning the intermediate 
q-values much more clearly and is closer to a correct identification of the notion of a path-integral.

Finally, the third edition of the book \cite{diracbook3}. Again, Dirac makes major changes in his presentation. He decides to present 
both the proofs of the analogy i.e the one presented in the paper but which was absent in the second edition, as well as the one he 
presented in the second edition of the book. But the analogy is presented now as an equality! He defines
\be
\label{eq:Stilde}
\langle\,q^\prime_t|q^{\prime\prime}_T\rangle\,
\equiv\,e^{\frac{i\,{\tilde S}}{\hbar}}
\ee
The ${\tilde S}$ is in general \emph{complex}.
This way Dirac makes the earlier analogy into an equality in the $\hbar\,\rightarrow\,0$ limit with the possibility of an overall 
proportionality. The treatment of the issues pertaining to the path-integral interpretation remain the same as in the second edition. 

It is very important to emphasize that this equality, for short time separations, would become
\be
\label{eq:diracmissing}
\langle\,q^\prime_{t+\epsilon}|q^{\prime}_t\rangle\,
=\,e^{\frac{i\,{\tilde S}}{\hbar}}
\ee
This is also a \emph{missing link}, very different from that of Feynman as expressed in eqn.(\ref{eq:missinglink0}). Therefore, one ought
to be able to construct completely different classs of path-integrals starting from eqn.(\ref{eq:diracmissing}). That is the main thrust
of this work.
The fourth edition of the
book 1958 \cite{diracbook4}, as well as the revised fourth edition \cite{diracbook4r} were identical to the third edition, as for as the
issues of this paper are concerned. So it is reasonable to take these as representing Dirac's final thoughts on the matter.

\subsection{Quantisation ambiguities and the Feynman path-integral}
It is well known that going from classical mechanics to quantum mechanics is in general not expected to be unique. This should hardly come as a
surprise since the classical analogue can be thought of as the $\hbar\,=\,0$ case and knowing  $f(0)$ there is no unique way of fixing
the $f(x)$! Feynman was of course aware of it, and we have quoted his views explicitly on p.5-6. There are essentially \emph{two} sources
of such non-uniqueness. One 
is the ordering ambiguity. In classical theory, momentum p and position q, being {\it commutative} can be written in any
order i.e qp and pq are one and the same. On the other hand, in quantum theory, the corresponding {\it operators} ${\hat q},\,{\hat p}$
do not commute
\be
\label{eq:heisenbergalgebra}
[{\hat q},\,{\hat p}]\,=\,i\hbar
\ee
Consequently, for example, the candidate Hamiltonians (for the moment we take any Hermitian operator as a possible Hamiltonian, ignoring
issues of whether they are bounded from below) ${\hat p}\,{\hat q}\,{\hat q}\,{\hat p}\,, {\hat p}\,{\hat p}\,{\hat q}\,{\hat q}\,$ which are 
obviously different quantum mechanically have the same classical analogue $p^2q^2$.  The other source of ambiguities, in a sense, are 
explicitly $\hbar$-dependent terms in the quantum Hamiltonians like $\hbar\,{\hat{\tilde H}}$
which vanish in the classical limit understood to be the $\hbar\,\rightarrow\,0$ limit. 

Either way, every choice of the quantum Hamiltonian results in general in a different
quantum description. Specifically, they lead to different observable consequences. The eigenvalues, wavefunctions, probability amplitudes
will be generically different. Therefore, $\langle q_{t+\epsilon}^\prime|q_t^\prime\rangle$ will also be different. These are consequences
of the rules for quantum theory, and are logically independent of any considerations based on path-integrals. 

Even before considering the implications of these ambiguities for path-integrals, it is important to look at their impact on the \emph{missing link} of eqn.(\ref{eq:missinglink0}). For reasons just explained, the l.h.s of this eqn  takes different values. But all these different
quantum Hamiltonians have the \emph{same} classical limit, and hence the same $S_{cl}$. Consequently, they will have the same value for
the r.h.s of eqn.(\ref{eq:missinglink0})! 

So there is a clear inconsistency, in general, between Feynman's  missing link relationsand non-uniqueness of quantisations of classical 
systems. 
The Feynman missing link relation can be correct for at best one class of quantisations. 
The resolution of this inconsistency has to lie in that eqn.(\ref{eq:missinglink0}) can not be true in general but only holds for certain 
classes of quantisations.  There is an interesting anecdote narrated on p.226 of \cite{gleick}. "Now he asked Dirac whether the great man 
had known all along that the
two quantities were proportional. "Are they?" Dirac said. Feynman said yes, they are. After a silence he (Dirac) walked away". Contrary to
what Gleick and numerous others, Feynman included, had stated, that the analogue was, in fact, exactly proportional, we have argued here 
how they can not be, in general. In fact Gleick goes so far as to assert "There was a rigorous and potentially useful mathematical bond".
This encounter, according to Gleick is supposed to have taken place in 1947, during the bi-centennial of the founding of Princeton university.
In my opinion, Dirac's response probably meant that Dirac was not sympathetic to the claims of their equivalence. That Dirac, as late as 1967,
twenty years after this alleged anecdote, maintained his analogue description \cite{diracbook4r} is added support to this opinion!

Let us
examine carefully how Feynman arrived at eqn.(\ref{eq:missinglink0}).
He explicitly showed that for Lagrangians of the type
\be
\label{eq:quadL}
L(x,{\dot x})\,=\,\frac{1}{2}\,M\,{\dot x}^2\,-\,V(x)
\ee
this eqn. reproduced the Schr\"odinger equation with the quantum Hamiltonian 
\be
\label{eq:quadH}
{\hat H({\hat x},{\hat p})}\,=\,\frac{{\hat p}^2}{2\,m}\,+\,V({\hat x})
\ee
To arrive at this, Feynman had to fix the normalisation $A(\epsilon)$ to be
\be
\label{eq:normfeyn}
A(\epsilon)\,=\,\sqrt{\frac{2\,\pi\,i\,\epsilon\,\hbar}{m}}
\ee
This is indeed a special case in that the quantum Hamiltonian is {\it insensitive} to ordering! It could be that the Feynman missing link
relation is consistent
only for such Hamiltonians which do not have a factoring ambiguity. But even for this special class, the second source of ambiguity is
still there, so the situation is not entirely clear. Differently stated, the Feynman missing link picks out one of the
(infinitely) many possible quantisations of a classical system. 

Whatever was said about the missing link can be immediately to generalized to the Feynman path-integral of eqn.(\ref{eq:feynpathint0}).
Now, the l.h.s, the Kernel $\langle q_T^\prime|q_t^\prime\rangle$ will generically take on different values for different quantum Hamiltonians,
albeit with the same classical limit, while the r.h.s is determined entirely by that unique classical limit. The inconsistency manifests
again.

This immediately raises a serious issue: if there is no unique path-integral representation, how can one understand the least action principle
of classical mechanics? In the Feynman path-integral, the classical action appears explicitly, and applying Dirac's stationary phase ideas
immediately gives rise to the least action principle.

A related issue is, what are the implications of using the eqn.(\ref{eq:diracmissing}) instead of eqn.(\ref{eq:missinglink0}) to
construct the path-integrals? Since ${\tilde S}$ need not be uniquely determined by $S_{cl}$, there is at least some hope to remove
the inconsistency forced on by the latter eqn. In other words, different choices of ${\tilde S}$ may take care of the quantisation ambiguities.

In the remainder of this paper, we shall show how Dirac tackles this central question by carefully analysing his paper and the different editions of his book. In fact he comes up with a path-integral far more general than Feynman's, completely consistent with quantisation ambiguities and yet
yields the least action principle of classical mechanics. 

\section{Dirac and the path-integral}
Here we carefully examine \cite{diracpaper, diracbook2, diracbook4r} to give support to the claims made earlier about Dirac's contributions
to the path-integrals. We will also answer Feynman's questions {\bf\,FQ1,\,FQ2}. We will chiefly look at i) proofs of the analogies, ii) 
path-integrals and quantisation ambiguities, and, iii) least action principle in classical mechanics, and, iv) the new path integrals and the
Schr\"odinger eqn. 
\subsection{Classical canonical transformations}
As a prelude to Dirac's proofs of
the analogies, we begin by reviewing classical canonical transformations.  
These are transformations that take an old pair of canonical variables $q,p$ to a new pair $Q,P$. In other words
\be
\label{eq:canonicaltr}
\{q\,,p\}_{PB}\,= 1; \{Q\,,P\}_{PB}\,=\,1
\ee
where 'PB' stands for {\it Poisson Brackect}. All other PB's are zero. Recall that the so called {\it contact} transformations given by
$Q\,=\,Q(q,t)$ leave the Euler-Lagrange equations of motion unchanged. Furthermore, these are transformations taking place entirely on
the {\it configuration space}. 

On the other hand, Hamilton's equations enjoy a much wider freedom i.e all the canonical transformations of the type
\be
\label{eq:canonical2}
Q\,=\,Q(q,p,t)\quad\quad\quad\,P\,=\,P(q,p,t)
\ee
They have to however satisfy
\be
\label{eq:canonical3}
{\dot Q}\,=\,\frac{\partial\,H^\prime}{\partial\,P}\quad\quad\quad\,{\dot P}\,=\,-\frac{\partial\,H^\prime}{\partial\,Q}
\ee
where $H^\prime(Q,P)$ is the new Hamiltonian. As is well known, such canonical transformations can be described in terms of the so called
{\it Generating Functions} \cite{landau}. For the choice when the generating function S is a function of $(q\,,Q\,,t)$ one gets
\be\label{eq: generating}
\frac{\partial\,S}{\partial\,q}\,=\,p\quad\quad\quad \frac{\partial\,S}{\partial\,Q}\,=\,-\,P\quad\quad \frac{\partial\,S}{\partial\,t}\,=\,H^\prime\,-\,H
\ee
This is as far as canonical transformation at some time t are concerned. It is also known in classical mechanics that under time evolutions
the change in the dynamical variables (q,p) is also described by canonical transformations! If $q_t\,,p_t$ are the dynamical
variables at initial time t, and, $q_{t^\prime}\,,p_{t^\prime}$ are dynamical variables at final time $t^\prime$, the corresponding
canonical transformations obey
\be
\label{eq:canonicalevol}
p_t\,=\,\frac{\partial\,S}{\partial\,q_t}\quad\quad\,\frac{\partial\,S}{\partial\,t}\,=\,-\,H;\quad\quad p_{t^\prime}\,
=\,-\,\frac{\partial\,S}{\partial\,q_{t^\prime}}\quad\quad\,\frac{\partial\,S}{\partial\,t^\prime}\,=\,H
\ee
Integrating these equations, one can show that S is
\be
\label{eq:principal}
S\,=\,\int_t^{t^\prime}\,L\,
\ee
i.e S is the action(also Hamilton's first principal function).

\subsection{Quantum analogues}
Dirac then proceeds to prove the analogy for quantum mechanics. The proof given in the paper \cite{diracpaper} is as follows: he considers
the transformation function and defines(I have changed the notation to avoid confusion; this is the same as the earlier introduced 
eqn.(\ref{eq:Stilde}))
\be
\label{eq:qanalog}
\langle\,q_{t^\prime}|q_T^{\prime}\rangle\,
\equiv\, e^{i\,\frac{{\tilde S}}{\hbar}}
\ee
here primes and double-primes indicate eigenvalues of corresponding operators and the above equation {\it defines} ${\tilde S}$ which,
as already emphasized, has no a priori relation to the classical action, and in fact need not even be real. But the important point is 
that ${\tilde S}$ is an
\emph{ordinary function} and not an \emph{operator}.

To handle the non-commutativity inherent in quantum mechanics, Dirac introduces a very novel technique \cite{diracpaper}:for every operator
$\alpha$ he introduces a {\it mixed representation} $\langle\,q^\prime|\alpha|Q^\prime\rangle$. To reduce confusion the variables 
${\hat q}_t, {\hat q}_T$
have been renamed ${\hat q}\,,{\hat Q}$. The mixed representation can easily be related to the standard representation where matrix elements 
are evaluated 
in any basis. It is easy to see that
\be
\label{eq:mixedrep}
\langle\,q^\prime\,|f({\hat q})\,G({\hat Q})\,|Q^\prime\,\rangle\,=\,f(q^\prime)\,G(Q^\prime)\,\langle\,q^\prime\,|\,Q^\prime\rangle
\ee
Dirac then introduces the notion of {\it well-ordered} operators. They are operators of the form 
\be
\label{eq:wellordered}
\alpha(q\,,Q)\,=\,\sum_i\,f_i(q)\,G_i(Q)
\ee
By using the commutation rules of quantum mechanics every operator can be brought into a well-ordered form, though in practice it can be
quite unwieldy to do so. Given a pair of \emph{non-commuting} operators ${\hat q}, {\hat Q}$ the well-ordered form of every operator has
to be \emph{unique}. Else, the mixed-representation of that operator would not be unique, which in turn would imply the standard representation
in any basis would not be unique.

Then, on using eqn.(\ref{eq:mixedrep}) it is easy to see that
\be
\label{eq:mixedrep2}
\langle\,q^\prime\,|\alpha(q\,,Q)\,|\,Q^\prime\,\rangle\,=\,\alpha(q^\prime\,Q^\prime\,)\,\langle\,q^\prime\,|\,Q^\prime\,\rangle
\ee
Applying to the momentum operators $p,P$ given by
\be
\label{eq:momenta}
\langle\,q^\prime\,|p\,=\,-\,i\hbar\,\frac{\partial}{\partial\,q^\prime}\,\langle\,q^\prime\,|\quad\quad\quad
P\,|\,Q^\prime\rangle\,=\,i\hbar\,\frac{\partial}{\partial\,Q^\prime}\,|\,Q^\prime\rangle
\ee
On using
\be
\langle q^\prime|Q^\prime \rangle\,\equiv\,e^{i\,\frac{{\tilde S}(q^\prime,Q^\prime)}{\hbar}}
\ee
Dirac shows that
\be
\label{eq:qanalog2}
p(q^\prime\,,Q^\prime)\,=\,\frac{\partial\,{\tilde S}(q^\prime,Q^\prime)}{\partial\,q^\prime}\quad\quad\quad
P(q^\prime\,,Q^\prime)\,=\,-\,\frac{\partial\,{\tilde S}(q^\prime,Q^\prime)}{\partial\,Q^\prime}
\ee
These are exactly of the same form as eqns.(\ref{eq:canonicalevol}), except for the time-derivative parts, which we come to next.

The paper \cite{diracpaper} appeared in 1932. By that time the first edition of Dirac's book (1930) \cite{diracbook1} had already
appeared and this material nor any discussion of the least action principle was in it. In 1935 Dirac came out with the second edition
of the book \cite{diracbook2} with substantial revisions. Now we take up his proof of the quantum analogue in the second edition.

Dirac completely omits the discussion of the approach based on well-ordered operators. Instead, he provides the following differential equations
obtainable from Schr\"odinger equation:
\be
\label{eq:qanalog2ed}
i\,\hbar\,\frac{d}{dt}\,\langle\,q^\prime\,|\,Q^\prime\rangle\,=\,\int\,\,\langle\,q^\prime\,|\,H\,|\,q^{\prime\prime}\,\rangle\,q^{\prime\prime}\,
dq^{\prime\prime}\,\,\langle\,q^{\prime\prime}\,|\,Q^\prime\rangle
\ee
and, likewise
\be
\label{eq:qanalog2ed2}
-\,i\,\hbar\,\frac{d}{dT}\,\langle\,q^\prime\,|\,Q^\prime\rangle\,=\,\int\,\,
\,\langle\,q^{\prime}\,|\,Q^{\prime\prime}\rangle
dQ^{\prime\prime}\,
\langle\,Q^{\prime\prime}\,|\,H\,|\,Q^\prime\,\rangle\,
\ee
On applying the well-ordered operator method, one can obtain the temporal differential equations
\be
\label{eq:qanalog2edtemp}
\frac{\partial\,{\tilde S}(q^\prime,Q^\prime)}{\partial\,t}\,=\,-\,H(q^\prime,Q^\prime)\quad\quad\quad\,
\frac{\partial\,{\tilde S}(q^\prime,Q^\prime)}{\partial\,T}\,=\,\,H(q^\prime,Q^\prime)
\ee
It is critical to have these in order to integrate them to get the analog of eqn.(\ref{eq:principal}). But the equations in second edition
do not seem to transparently lead to eqn.(\ref{eq:qanalog2})! It is to be appreciated that the eqns.(\ref{eq:qanalog2,qanalog2edtemp})
involve only commuting objects with no explicit trace of the non-commutativity of quantum mechanics. 

That brings us to the third edition \cite{diracbook3}.
In the third edition, Dirac reinstates the proof given in the paper along with the equations of the second edition. Thus the proof of the 
analogy is mathematically precise. {\bf It holds for all values of $\hbar\,,t\,,T$}. There is no question of the analogues being either
equal or even proportional, even for short temporal separations. A good example in this context are \emph{Poisson Brackets} in classical
mechanics and \emph{commutators} in quantum mechanics. They are analogues of each other in the sense of obeying the same algebra, but
they arev neither equal, nor proportional to each other.

That should answer Feynman's first question, FQ1, emphatically in the negative.
\subsection{Dirac Path-integrals}
Dirac sets out to investigate the consistency of the analogy
\be
\label{eq:consistency}
B(t_1,t_2)\,=\,e^{i\,\frac{S_{cl}}{\hbar}}\,=\,e^{\frac{i}{\hbar}\,\int_{t_1}^{t_2}\,L}\,\sim\,\langle\,q^\prime\,|\,Q^\prime\rangle
\ee
where we have used the symbol $\sim$ to denote analogy. Before proceding, it is very important to note that, for infinitesimal temporal i
separations $B(t,t+\epsilon)$ only depends on the q-values at the two ends. Dirac investigates the consistency by dividing the interval
$[t_i\,,\,t_f]$ into a very large number of intervals with the intervening times labelled as $t_m$(note that once again a change in notation
has been made whereby $t_i\,=\,t;t_f\,=\,T$! On the classical side one has
\be
\label{eq:consistency1}
B(t_i\,,\,t_f)\,=\,B(t_i,t_1)\,B(t_1,t_2)\,\ldots\,B(t_m\,,\,t_{m+1}\,\ldots\,B(t_N,t_f)\,=\,e^{\frac{i}{\hbar}\,\int_{t_i}^{t_f}\,L}
\ee
It is very important to bear in mind that $B(t_i,t_f)$ depends, in addition to the positions at the initial and final times, on all the 
intermediate positions $q_m$ at $t_m$ and that these positions lie on the classical trajectory connecting the initial and final positions.
Now Dirac writes down the quantum analogue of the above
\be
\label{eq:qconsistency}
\langle\,q_{t_f}^\prime\,|\,q_{t_i}^\prime\,\rangle\,=\,\int\ldots\int\,\langle\,q_{t_f}^\prime|q_N^\prime\rangle\,dq_N^\prime\ldots
dq_1^\prime\,\langle\,q_1^\prime|q_{t_i}^\prime\rangle
\ee
All that has been used is the completeness for all the intermediate states. On using the {\it definition} of ${\tilde S}$ this becomes
\be
\label{eq:diracpathint}
\langle\,q_{t_f}^\prime\,|\,q_{t_i}^\prime\,\rangle\,=\,\int\,\ldots\int\,\prod_{i=1}^{N}\,dq_i^\prime\,e^{\frac{i}{\hbar}\,{\tilde S}(\{q_i^\prime\})}
\ee 
where we have used the shorthand notation $\{q_i^\prime\,\}$ to denote the initial, final as well as all the intermediate positions. This
has been delivered entirely by the analogy and is valid for {\it arbitrary} $\hbar$. This is what we like to call the {\bf Dirac Path-integral}.It is for certain a path-integral.

At first sight it might appear that a major weakness of this path-integral is that nothing much seems to be known about ${\tilde S}$ making
 eqn.(\ref{eq:diracpathint}) not particularly useful. We shall now show that that is actually its major strength for it allows this
{\it generalized} path-integral to be fully compatible with quantisation ambiguities which the Feynman path integral could never be.

Though the analogy is unable to fix ${\tilde S}$ there is much that can nevertheless be said about it, as Dirac has already done. In 
particular, as we approach $\hbar\,\rightarrow\,0$ ${\tilde S}$ must increasingly {\it equal} the classical action.  That is,
\be
\label{eq:tildeS}
\hbar\,\rightarrow\,0\,\rightarrow\,{\tilde S}\,\rightarrow\,S_{cl}
\ee
We can reexpress this in a different way:if we write
\be
\label{eq:starS}
{\tilde S}\,=\,S_{cl}\,+\,S^*(\hbar)
\ee
Then 
\be
\label{eq:starS2}
S^*\,\rightarrow\, S^*_{norm} \quad\quad\quad as\quad\quad \hbar\,\rightarrow 0
\ee
and the final form of the Dirac path-integral would be
\be
\label{eq:diracpathint*}
\langle\,q_{t_f}^\prime\,|\,q_{t_i}^\prime\,\rangle\,=
\,\int\,\ldots\int\,\prod_{i=1}^{N}\,dq_i^\prime\,e^{\frac{i}{\hbar}\,S_{cl}\,+\,S^*}
\ee 
Two very important features of this path-integral: though it is valid for all $\hbar$, for very small $\hbar$, the $S_{cl}$ dominates (note
that it is not the only term in this limit) and Dirac's stationary phase arguments immediately lead to the least action principle for classical
analogue.

The other is that away from $\hbar\,\rightarrow\,0$, the $S^*$ terms are indeed important, and there is no more any conflict between the
path-integral and the quantisation ambiguities; both the LHS and RHS are sensitive to these ambiguities and in fact $S^*$ can be taken as a
way of classifying the ambiguities. This is in stark contrast with the Feynman path-integral for which $S^*\,=\,S^*_{feyn}$. On using
eqn.(\ref{eq:missinglink0}), it is easy to see that
\be
\label{eq:S*feyn} 
S^*_{feyn}\,=\,i\,\hbar\,\ln{A}\,=\,\frac{i\,\hbar}{2}\,\{\,\ln{\frac{2\,\pi\,i\,\hbar}{m}}\,+\,\ln{\epsilon}\}
\ee
This does satisfy $S^*_{feyn}\,\rightarrow\,0$ as $\hbar\,\rightarrow\,0$ as required by eqn.(\ref{eq:starS2}).

$S^*$ need not be equal to $S^*_{feyn}$ in general, but that choice can be taken as a particular choice of quantisation for which the 
Feynman path-integral is the
correct representation. In the absence of a deeper guiding principle for quantum Hamiltonians, any such choice is ad hoc in itself.
Only empirical vindications can be used to validate the quantum Hamiltonians.
\section{Dirac Path Integal and Schr\"odinger Equation}
Feynman justified his eqn.(\ref{eq:missinglink0}) by the fact that it led to the Schr\"odinger eqn. with what appeared to be the correct
quantum Hamiltonian, though, as mentioned earlier, there are no first principles criteria for determining quantum Hamiltonians. In this section
we show how Dirac path-integrals too generically lead to Schr\"odinger eqn. We illustrate the connection between Dirac path integrals and 
Schr\"odinger eqn with the help of an explicit example. We consider the 
{\it short-time} propagator
\be
\langle x,t+\epsilon|y,t\rangle
\ee
of a quantum theory whose {\it Quantum Hamiltonian} will be determined in accordance with the choice(just for illustration) 
\be
\label{eq:Sstar}
S^*(x,y,\epsilon)\,=\,\lambda\,\hbar\,\epsilon\,x^2\,{\hat x}^2\,+\,i\,\hbar\ln{A(x,\epsilon)}
\ee
We have allowed for the \emph{normalisation factor} to also depend on x. Furthermore, we take the classical Lagrangean to be
\be
L(x, {\dot x})\,=\,\frac{m}{2}\,{\dot x}^2\,-\,V(x)
\ee
Following Feynman, the wavefunctions $\psi(y,t),\psi(x,t+\epsilon)$ can be related(this is a general property of quantum mechanical wave
functions)
\be
\psi(x,t+\epsilon)\,=\,\int\,\frac{dy}{A(x)}\,e^{\frac{i\epsilon}{\hbar}\,\{\frac{m}{2}\,\frac{(x-y)^2}{\epsilon^2}\,-\,V(x)\,+\,\lambda\hbar\,x^2\,\frac{(x-y)^2}{\epsilon^2}\}}\,\psi(y,t)
\ee
Again, following Feynman, one writes $y\,=\,x\,+\eta$,and, introducing 
\be
f_0(x)\,=\,m\,+\,2\,\lambda\,\hbar\,x^2
\ee
the integral equation takes the form
\be
\psi(x,t+\epsilon)\,=\,\int\,\frac{d\eta}{A(x)}\,\,e^{\frac{i}{\hbar}\{f_0(x)\,\frac{\eta^2}{2\epsilon} \,-\,\epsilon\,V(x)\}}\,
\,\psi(x+\eta,t)
\ee
Note that this is exactly of the same form as the integral eqn in Feynman's case except for the replacements
\be
m\,\rightarrow\,f_0(x)\quad\quad\quad\,A\,\rightarrow\,A(x)
\ee
Due to the $\frac{\eta^2}{\epsilon}$ factors in the exponent, as 
$\epsilon\rightarrow\,0$, only very small $\eta$ contributions dominate the integral. Simply mimicking all the steps followed by Feynman, one 
ends up with the Schr\"odinger eqn0(x)\,\
\be
i\,\hbar\,\frac{\partial\,\psi(x,t)}{\partial\,t}\,=\,-\,\frac{\hbar^2}{4}\,
\{\frac{1}{f_0}\,\frac{{\partial}^2\,\psi(x,t)}{{\partial}^2\,x}\,+\, \frac{{\partial}^2\,\psi(x,t)}{{\partial}^2\,x}\,\frac{1}{f_0}\}
\,+\,V(x)\psi(x,t)
\ee
In other words, the choice of $S^*$ as per eqn.(\ref{eq:Sstar}) leads to the quantum Hamiltonian
\be
\label{eq:HqSstar}
{\hat H}_{\lambda}\,=\,\frac{1}{4}\,\{\,{\hat p}^2\,\frac{1}{f_0({\hat x}}\,+\,\frac{1}{f_0({\hat x}}\,\}\,+\,V({\hat x})
\ee
The following are to be noted:
\begin{itemize}
\item The Feynman case is $\lambda\,=\,0$. For this choice
\be
\label{Hqfeyn}
{\hat H}\,=\,\frac{{\hat p}^2}{2\,m}\,+\,V({\hat x})
\ee
\item In general, the Dirac path-integrals, classified by the choice of $S^*$, still yield the Schr\"odinger eqn, but with quantum 
Hamiltonians given by eqn.(\ref{eq:HqSstar}), generically different from the Feynman case.
\item In the classical limit, understood as the limit $\hbar\,\rightarrow\,0$, all the quantum Hamiltonians take the same limiting
\be
\label{eq:Hcl}
{\hat H}_{\lambda}\,\xrightarrow{\hbar\rightarrow\,0}\,H_{cl}\,=\,\frac{p^2}{2m}\,+\,V(x)\,\,\forall\,\lambda
\ee
\item The $S^*$ for the Feynman case($\lambda\,=\,0$) is {\bf not} $S^*\,=\,0$, but
\be
S^*_{feyn}\,=\,\ln{A}\quad\quad A\,=\,\sqrt{\frac{2\,\pi\,i\,\hbar\,\epsilon}{m}}
\ee
\item This simple example illustrates all the main points of the paper.
\end{itemize}
\section{Conceptual highlights of path-integral formulation of QM}
\label{sec:highlights}
In this section we address some important conceptual highlights of the path-integral formulations. The massive impact of this formulation 
on a variety of
areas, including financial markets \cite{kleinert} will not be our focus; instead, we shall focus on the conceptual aspects. 
This section should also enable readers to make a comparison between the Feynman path-integrals of eqn.(\ref{eq:feynpathint0}) and the newly
introduced Dirac path-integrals of eqn.(\ref{eq:diracpathint*}). 
We shall restrict ourselves to particles in one spatial dimension. 
\subsection{QM entirely in terms of {\it commuting} variables}
The LHS of either of these equations is a purely quantum mechanical object, calculable entirely within the quantum mechanical formalism,
which necessarily involves \emph{non-commutativity}. But the RHS is composed of a classical action $S_{cl}$ in the Feynman case, and,
$S_{cl}\,+\,S^*$ in the Dirac case,  along with all
the intermediate 'classical positions' as well as integrations over them. This is nothing short of a {\it miracle} as all traces of operators,
non-commutativity etc have disappeared in this representation of Quantum Mechanics! Nevertheless, it is completely equivalent to the
traditional formulation of quantum mechanics. This magic is entirely due to Dirac and his use of well-ordered operators to completely
obviate the explicit use of non-commutativity. Since Feynman's original construction was based on Dirac's analogy, which directly resulted
from the well-ordered techniques, this particular magic in Feynman path-integrals is a reflection of the magic of Dirac's techniques.

As is well known, Feynman showed that even the manifestly non-commutative relation
\be
\label{eq:commutation}
[{\hat x},{\hat p}]\,=\,i\,\hbar
\ee
emerges from the purely commutative calculus of path-integrals!
Both Feynman's thesis \cite{thesis} as well as his paper \cite{feynpaper} are a treasure-house for a
deeper understanding of the path-integrals. The apparent mystery gets resolved on noting that in the path-integral, momentum does not
appear explicitly and all the subtelety resides in a careful construction of it.

We have not demonstrated this in our paper. But we give a fairly good argument for why it should follow from Dirac path-integrals
as well. The key observation is that eqn.(\ref{eq:commutation}) does not depend on the specific form of $S_{cl}$. For the purposes
of this demonstration,one can treat the entire $S_{cl}\,+\,S^*$ of Dirac path-integrals as an effective $S_{cl}$. Then Feynman's
demonstration of eqn.(\ref{eq:commutation}) can be extended to the Dirac case as well. Of course, this effective can not be a true
classical action as it will explicitly involve $\hbar$.
 
This is an apt place to make some remarks about various alternative approaches to path-integrals and how non-commutativity issues are
handled by them. After Feynman's construction, several approaches to deriving the path integral have appeared in the literature. One
of these is to consider the quantum Hamiltonians of the type given in eqn.(\ref{eq:quadH}). The terms $T({\hat p}),V({\hat x})$
do not commute with each other. Working in the Schr\"odinger picture, the \emph{time evolution} operator $e^{i\frac{{\hat H}{\bar T}}{\hbar}}$
(without loss of generality we set $t\,=\,0$ and denote final time as ${\bar T}$ to avoid confusion with the kinetic energy T) is rewritten as
\be
\label{eq:slicedTevol}
e^{i\frac{{\hat H}{\bar T}}{\hbar}}\,=\,\xrightarrow{n\,\rightarrow\,\infty}\,(e^{i\,\frac{{\hat T}\,\epsilon}{\hbar}}\cdot\,
e^{i\,\frac{{\hat V}\epsilon}{\hbar}})^n
\ee
on using the famous {\bf Trotter's Formula}. One then cleverly uses completeness of both position bases and momentum bases to arrive
at the path-integral. But in Dirac's, and consequently Feynman's, Trotter's formula plays no role whatsoever, for reasons explained above. 
Furthermore, if the Hamiltonians
are not of the form of eqn.(\ref{eq:quadH}) it is not at all clear if Trotter's formula would be of any use.

In the \emph{quantum histories} approach \cite{griffiths,histories}, the summands are chains of quantum operators unlike both the Dirac 
and Feynman 
path-integrals, where the summands are complex numbers. It would be a very important exercise to understand the inter-relationships between
all these approaches. In particular, it will be interesting to explore how Dirac's well ordered-operators approach can be used in this context.

Lastly, we refer to the remarkable works of Gregor Wentzel \cite{wentzel1,wentzel2} which appeared even before the works of Heisenberg and 
Schr\"odinger. A good account of Wentzel's works, and their significance, can be found in \cite{antoci}. According to these authors, Wentzel,
in this important but forgotten work, identified $\int\,L\,dt$ as a phase that governs \emph{quantum interference}. I have not fully
digested Wentzel's ideas, but they have to be understood. Certainly from the point of view understanding the non-commutativity aspects as well
as those of quantisation ambiguities.

\subsection{Least action principle of classical mechanics}
In classical mechanics the {\it principle of least action}, namely, that the path followed by a system under dynamics is the one
that renders the action $S\,=\,\int\,L\,dt$(seen as a {\it functional} of paths) stationary i.e
\be
\label{eq:leastactioncm}
\delta\,S\,=\,0
\ee
appears from nowhere without any understanding of why it has to be so. Its main raison d'etre in classical mechanics is that it correctly
reproduces the equations of motion. The Lagrangian version of the action principle is
\be
\label{eq:eulerlagrange}
\delta\,\int\,L(q,{\dot q})\,dt\,=\,0\,\rightarrow\,\frac{\partial\,L}{\partial\,q}\,=\,\frac{d}{dt}\,\frac{\partial\,L}{\partial\,{\dot q}}
\ee
while the Hamiltonian version is
\be
\label{eq:hamiltoneqns}
\delta\,\int\,(p\,{\dot q}\,-\,H(p,q))\,dt\,=\,0\,\rightarrow\,\frac{\partial\,H}{\partial\,q}\,=\,-\,{\dot p}\,;\,\frac{\partial\,H}{\partial\,p}\,=\,{\dot q}
\ee
The least action principle appears more as a pnemonic, without any deeper physical significance. The path-integral formulation is supposed
to provide that significance by showing how it emerges 'naturally from the path-integral in the $\hbar\,\rightarrow\,0$ limit. To get the
crux of the argument(first given by Dirac in \cite{diracpaper}), consider fixing all intermediate $x^\prime$'s except one, say,
$x_k^\prime$(in the Dirac path-integral of eqn.(\ref{eq:diracpathintegral*}) the notation used is $q_k^\prime$). 
Any variation,
including infinitesimal ones,
in $x_k^\prime$ will cause the phase $e^{\frac{iS_{cl}}{\hbar}}$ to fluctuate very rapidly in this limit(frequency approaching $\infty$), 
averaging to zero.
The only exception to this behaviour would be if $x_k^\prime$ were the stationary point of $S_{cl}$. Then for sufficiently small variations 
in $x_k^\prime$ 
$S_{cl}$ will not change, and all these nearby paths will make contributions to the path-integral that will constructively interfere. 
Of course, for
variations that are not that small, the earlier argument of destructive interference applies. As emphasized by Dirac, this line of
reasoning can be applied to all the intermediate $x^\prime$'s with the result that in this limit, the only path that contributes is the one that
extremizes the classical action. Feynman was so struck by the beauty of Dirac's reasoning that he chose to review the salient aspects in
his paper \cite{feynpaper}.

Now let us see how the same arguments can be applied to the Dirac path-integral of eqn.(\ref{eq:diracpathint*}). In the limit 
$\hbar\,\rightarrow\, 0$, the $S^*$ is \emph{sub-dominant} compared to $S_{cl}$, so the stationary paths, even for the Dirac case, are i
determined by $S_{cl}$ only!
\subsection{Connections to Classical Statistical Mechanics}
The RHS of the path-integrals either \ref{eq:feynpathint0}, or, eqn.(\ref{eq:diracpathint*}), formally look very much like 
the {\it partition function} of a 1-d classical
statistical system except for one crucial difference;the integrand of the path-integrals is a {\it complex phase} whereas the integrand of
a thermal partition function is of the form $e^{-\,\frac{H}{kT}}$. This suggests working out the quantum problem in {\it Euclidean time}. It
amounts to a {\it Wick rotation} $t\,\rightarrow\,i\,\tau$ where $\tau$ is {\it Euclidean time}. This in turn leads to $S\,\rightarrow i\,S_E$,
where $S_E$ is the {\it Euclidean Action}.
It turns out that for most systems be it 
quantum mechanics or Quantum Field Theories, the  Euclidean Action is {\it positive}(gravity is an exception). Thus we get the map
\be
\label{eq:euclidean}
e^{i\,\frac{S}{\hbar}}\,\rightarrow\,e^{-\,\frac{S_E}{\hbar}}\,\equiv\,e^{-\,\frac{H}{kT}}
\ee
Thus $\hbar$ plays the role of temperature and $S_E$ the role of the Hamiltonian for the equivalent classical statistical system, necessarily
in one higher spatial dimension!

This very deep connection could only be established in the path-integral approach. From a foundational point of view, this is the only
{\it constructive} approach to quantum field theories at present. By 'constructive' one means that all mathematical manipulations are
{\it well defined}! Strictly speaking, the space-time continuum has to be approximated by discrete set of points. This is in sharp contrast 
to Quantum Electrodynamics of Dirac, Feynman, Schwinger, Tomonaga and Dyson where almost 
every manipulation leads to {\it infinities}. Dirac found the situation highly unsatisfactory and while stating "... it is this high 
energy part that is
responsible for the infinities", goes on to add (section 81 {\it Interpretations} in the chapter on Quantum Electrodynamics in the revised
fourth edition of his book \cite{diracbook4r}) "It appears that we must modify the high energy part of the interaction. At present there 
does not 
exist any detailed theory of the other particles and so it is not possible to say how it ought to be modified. The best we can do is to cut it
out from the theory altogether, and so remove the infinities. The precise form of the cut-off and the energy where it is applied 
will be left unspecified. Of course, the cut-off spoils the relativistic invariance of the theory. This is a blemish which can not be avoided
in our present state of ignorance of high-energy processes."

This was said in 1967 and subsequently there have been such high-energy modifications to QED like Quantum Chromodynamics and the Standard
model of electroweak interactions. While these have ameliorated many of the defects of QED, they too are beset with infinities and the
high energy modifications have not fixed the blemishes alluded to by Dirac. Though Superstring theories are claimed to be {\it finite}, we
shall not comment on them here.

This equivalence of quantum field theories on a Lattice to classical statistical mechanics has had great impact and applicability. Chief 
among them is
{\it Lattice Gauge Theory} \cite{seiler} for QCD. The classical statistical mechanics systems can be numerically simulated relying on numerous techniques
from statistical mechanics. There has been explosive growth in this area \cite{creutz}. These formulations are of the cut-off type, with
the cut-off operating at the level of the entire theory instead of process by process. The loss of relativistic invariance by discretisation
is obvious.
\section{Conclusions}
In conclusion, we list some of the salient implications of this work.  We have shown that Feynman's  so called 'proof' that what Dirac had
thought was an analogy was in fact a proportionality, can not always be correct. It is merely a reflection of the quantisation ambiguities.That 
there could not have been any {\it general} method for proving Feynman's claim, and that there could not have been any {\it rigorous 
and mathematically useful bond} as claimed by his biographer Gleick \cite{gleick}.
In fact, Feynman had to base his assertions on very specific classes of Hamiltonians of the form $H\,=\,T(p)\,+\,V(q)$. He was aware of the
limitations and to quote him "....only proved in a very special case. It is apparent, however, that the proof may be readily extended to
any Lagrangian which is a quadratic function of velocities..". Even the said extension to a more general class can not be general enough to
be valid always.

This answers Feynman's first question FQ1 decidedly in the negative. His second question FQ2, was about the usefulness
of the analogues Dirac found. It is quite clear how decisive the analogy is in arriving at Dirac's path-integral for general quantisations.
That ought to be enough of an answer to FQ2. It can be dangerous to judge a scientific concept or even a notion
based on {\it usefulness}. A case in point is that even Feynman's path-integral found to be so highly useful in a wide variety of circumstances,
took several decades since its inception to solve the Hydrogen atom problem! 

But the really major results of this paper are i) Dirac's contributions were not restricted to revealing the role of Lagrangian in
quantum mechanics, he in fact discovered the path integral, ii) giving an explicit equation(eqn.(\ref{eq:diracpathint*}) for this 
path-integral, which we have named \emph{the Dirac path integral},iii) the Feynman path integrals are inconsistent with quantisation ambiguities
inherent in quantum mechanics, iv) in contrast, the Dirac path integrals are fully consistent with such ambiguities, v) we have shown
how the large class of Dirac path-integrals account for the principle of least action of classical mechanics.

What gave Feynman confidence in his eqn.(\ref{eq:missinglink0}) was that it reproduced the Schr\"odinger eqn. with one particular quantum
Hamiltonian. With an explicit example, we
have shown that this is true of the corresponding eqn.(\ref{eq:diracmissing}) for the Dirac case too. But each choice of $S^*$ in the
Dirac case yields a different \emph{quantum Hamiltonian}. We have also shown that all those quantum Hamiltonians have the \emph{same}
classical limit. So one of the implications of our Dirac path-integral is as a systematic generator of quantum Hamiltonians with the
same classical limit.

We have carefully compared the way Dirac avoids the explicit use of non-commuting objects with various other approaches to path-integrals.
In particular we have stressed how Trotter's formula plays no role in Dirac's analysis, and how Dirac's \emph{well-ordered operator} method
fully handles all aspects of non-commutativity in terms of the commuting \emph{mixed-representation} of such operators. We have, at the same 
time contrasted the non-commutativity issues in approaches like the \emph{quantum histories} approaches \cite{griffiths}, and the 
Georgiev-Cohen \cite{histories} approaches.. The connections between these
approaches and the Dirac path-integrals, as well as the possible uses of Dirac's techniques in them are worthy of serious study.Likewise, we
have briefly discussed Wentzel's forgotten approach to quantum mechanics that even predated Heisenberg and Schr\"odinger but rather
remarkably uncovered the classical action as the phase for quantum interference. 
\acknowledgments I thank Danko Georgiev for many discussions.
Thanks are also to the Librarian of Raman Research Institute, Y.M. Patil, for making available the relevant
chapters from all the four editions of the book.I thank R. Ramachandran for bringing to my attention Wentzel's works and 
T.R. Govindarajan for refreshing my memory about the same. 


\begin{thebibliography}{}
\bibitem{kleinert} Hagen Kleinert, {\it Path Integrals in Quantum Mechanics, Statistics, Polymer Physics and Financial Markets}, World
Scientific Publishers, 2009.
\bibitem{seiler} Erhard Seiler, {\it Gauge Theories as a problem of Constructive Quantum Field Theory and Statistical Mechanics}, Lecture
Notes in Physics, Number 159, Springer-Verlag 1982.
\bibitem{gleick} James Gleick, {\it The Genius}
\bibitem{feynpaper} R.P. Feynman, {\it Space-time approach to non-relativistic quantum mechanics}, Reviews of Modern Physics, Vol 20, p.267
(1948)
\bibitem{nobel} Feynman's Nobel acceptance lecture (1965).
\bibitem{joking} R.P. Feynman, {\it Surely You are Joking, Mr Feynman!}, Vintage Publishers, 1992.
\bibitem{diracpaper} P.A.M. Dirac, {\it Lagrangian in Quantum Mechanics}, 
Physikalische Zeitschrift der Sowjetunion, Band 3, Heft 1 (1933),
\bibitem{thesis} L.M. Brown(Ed) {\it Feynman's Thesis - A New Approach to Quantum Theory}, World Scientific, 2005. This single
resource contains Feynman's Thesis {\it The Principle of Least Action in Quantum Mechanics}, his paper as well as Dirac's Lagrangian in Quantum Mechanics paper.
\bibitem{diracbook1} P. A. M. Dirac, The Principles of Quantum Mechanics (Clarendon Press, Oxford, 1st edn., 1930). 
\bibitem{diracbook2} P. A. M. Dirac, The Principles of Quantum Mechanics (Clarendon Press, Oxford, 2nd edn., 1935). 
\bibitem{diracbook3} P. A. M. Dirac, The Principles of Quantum Mechanics (Clarendon Press, Oxford, 3rd edn., 1947). 
\bibitem{diracbook4} P. A. M. Dirac, The Principles of Quantum Mechanics (Clarendon Press, Oxford, 4th edn., 1957). 
\bibitem{diracbook4r} P. A. M. Dirac, The Principles of Quantum Mechanics (Clarendon Press, Oxford, revised 4th edn., 1967). 
\bibitem{griffiths} R.B. Griffiths, {\it The Consistent Histories Approach to Quantum Mechanics}, in The Stanford Encyclopedia of
Philosophy (Summer 2019 Ed), Edward N Zalta(Ed.); https://Plato.Stanford.edu/archives/sum2019/entries/qm-consistent histories.
\bibitem{histories} D. Georgiev and E. Cohen, Physical Review A 97(5):052102, 2018.
\bibitem{wentzel1} G. Wentzel, {\it Zur Quantenoptik}, Zs.f.Physik,22,193-99, 1924.
\bibitem{wentzel2} G. Wentzel, {\it Zur Quantentheorie des R\"ontgenbremsspektrums}, Zs.f.Physik,27,237-284, 1924.
\bibitem{antoci} S. Antoci and Dierck-E. Liebscher, {\it The third way to quantum mechanics is the forgotten first}, arXiv:9704028v1[hist-ph].
\bibitem{creutz} M. Creutz, {\it Quarks Gluons and Lattices}, Cambridge Monographs on Mathematical Physics, 1985.
\bibitem{landau} Landau and Lifshitz, {\it Mechanics}.
\end{thebibliography}
\end{document}